\documentclass[12pt]{article}

\usepackage{graphicx}
\usepackage{amsmath}
\usepackage[numbers]{natbib}
\usepackage{url}
\usepackage{authblk}

\begin{document}

\title{Control of flow behavior in complex fluids using automatic differentiation}

\author[a,c]{Mohammed G. Alhashim \thanks{ mohammed.mga122@gmail.com} }
\author[b]{Kaylie Hausknecht }
\author[a,b]{Michael P. Brenner\thanks{ brenner@seas.harvard.edu}}

\affil[a]{School of Engineering and Applied Physics, Harvard University, Cambridge MA 02138}
\affil[b]{Department of Physics, Harvard University, Cambridge MA 02138}
\affil[c]{Saudi Aramco, Dhahran, Saudi Arabia 31311}



\maketitle

\begin{abstract}
Inverse design of complex flows is notoriously challenging because of the high cost of high dimensional optimization. Usually, optimization problems are either restricted to few control parameters, or adjoint-based approaches are used to convert the optimization problem into a boundary value problem. Here, we show that the recent advances in automatic differentiation (AD) provide a generic platform for solving inverse problems in complex fluids. To demonstrate the versatility of the approach, we solve an array of optimization problems related to active matter motion in Newtonian fluids, dispersion in structured porous media, and mixing in journal bearing. Each of these problems highlights the advantages of AD in ease of implementation and computational efficiency to solve high-dimensional optimization problems involving particle-laden flows.
\end{abstract}

Inverse particle laden flow design is a  PDE-constrained optimization problem that is important in diverse fields, ranging from tuning the rheology of suspensions, programming the self-assembly of colloidal particles, studying coalescence of emulsions, to tailoring the topology of porous media for filtration applications. 
The need to efficiently compute gradients with respect to high dimensional design variables has traditionally driven
 PDE-constrained optimization problems to be solved as boundary-value-problems using adjoint-based methods \cite{adjoint}. This approach has been successfully applied, for instance, to optimizing the shape of projectiles in finite Reynolds number flows \cite{marcusPF,shapeOPt} and  modulating microfluidic mixers \cite{mixingJFM}.

Analogous to adjoint-based methods, reverse mode automatic differentiation (AD) offers an efficient way of computing gradients with respect to design variables where the computational cost is independent of the number of control parameters. By applying the chain rule through a computation graph, AD circumvents the need to derive an explicit optimality condition \cite{autodiff}. AD has been a key driver of the machine learning revolution, enabling the development of deep neural network architectures that perform complex tasks that were unimaginable even a decade ago, ranging from large language models \cite{chatgpt} to generating artistic images based on text descriptions \cite{imageAI}. Despite the strong similarities between the 
mathematical structure of optimization problems for solutions of PDEs and machine learning, this method is less commonly used in fluid mechanics. Yet, it is straightforward to implement automatic differentiation in numerical solvers for fluid mechanics and can be carried out in open source machine learning libraries such as JAX \cite{jaxsoft}, TensorFlow \cite{tensorflow} and PyTorch \cite{pytorch}.

Here, we develop an end-to-end differentiable Fluid-Structure solver that implements the popular immersed boundary method \cite{peskin_2002,IBM_annual} for rigid bodies to address a class of inverse problems involving the intricate interplay between fluid dynamics and particle motion, such as the study of active matter propulsion, rheology of suspension system, etc. To showcase the versatility and power of the methodology, we tackle four different problems: (i) Optimizing the topology of a periodic structured porous medium to minimize the pressure drop in a pressure driven flow. (ii) Optimizing tracer dispersion in a periodic structured porous media where the tracer dispersion is modeled using Brownian dynamics. Solving this problem requires simultaneously optimizing over stochastic trajectories of the dispersed particles together with the underlying fluid mechanical optimization. (iii) Optimizing mixing in a 2-D journal bearing, where we focus on optimizing the design and protocol of rotating eccentric cylinders to maximize mixing efficiency,  showing the method easily applies to unsteady flows. (iv) Optimizing the propulsion efficiency of a swimmer. While the first three problems consider low Reynolds number flows, we explore the optimization of kinematic parameters to boost the efficiency of a cruising ellipse travelling at a constant velocity $U$ at Reynolds numbers in the range of $1000$. 

\begin{figure}[t]
\centering
\includegraphics[scale=0.35]{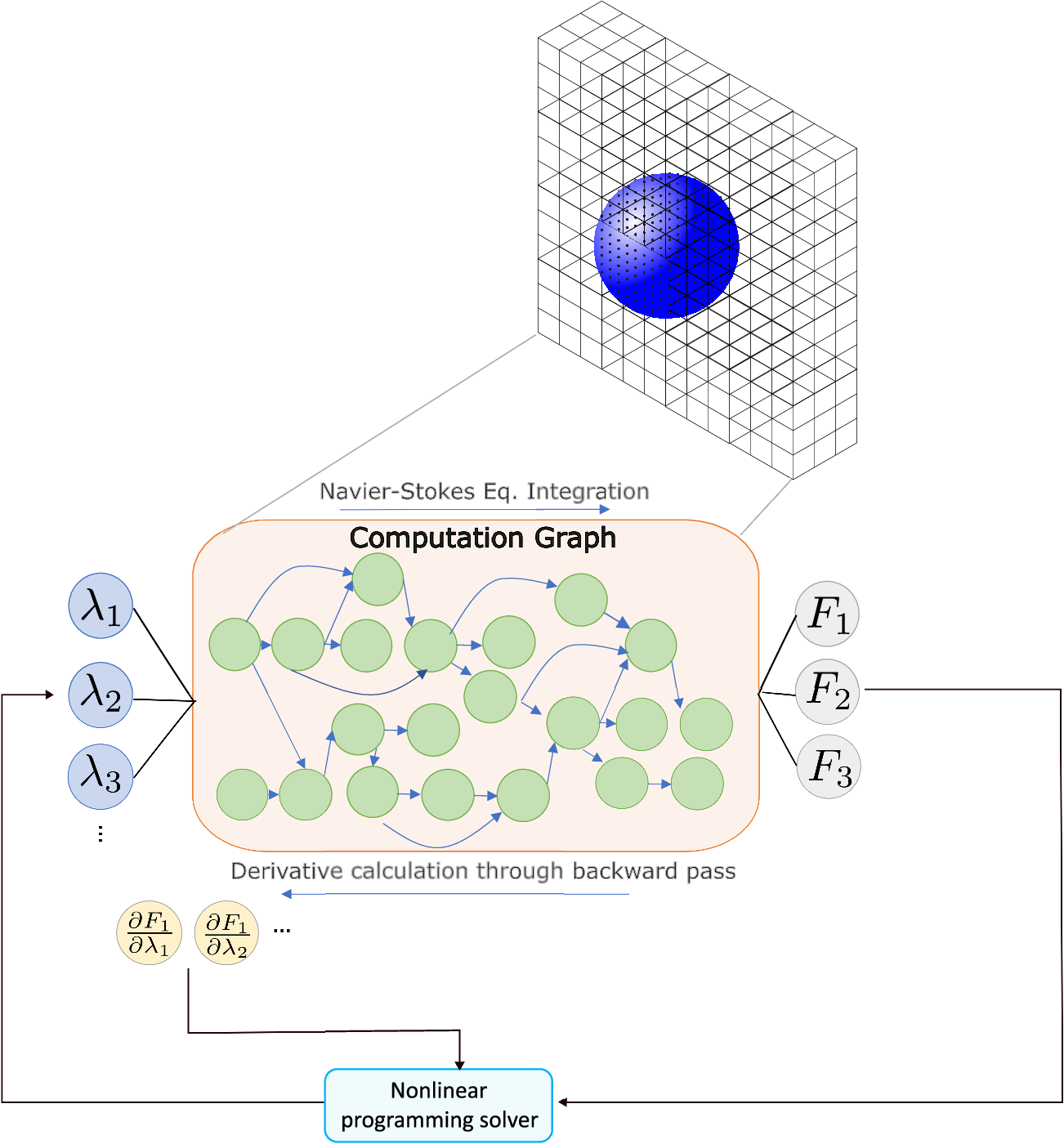}
\caption{\label{fig:chart} Flowchart of the optimization algorithm using a differentiable Immersed Boundary solver. The immersed boundary trajectory is unrolled with a computation graph. $F_i$, 
represent the loss function, equality and inequality constraints.}
\end{figure}

In each case, we compute the gradient of the loss function with respect to the optimization parameters using a unified computational framework (Figure \ref{fig:chart}), whereby the derivatives of the loss function are computed via a backward pass through the computation graph corresponding to the forward simulation. This is done without needing to derive or solve problem-specific adjoint equations. The versatility of this solver allows it to be seamlessly applied to various optimization challenges. This approach not only streamlines the optimization process but also opens up new possibilities for addressing complex fluid-structure interaction problems across various scientific and engineering disciplines. 

\section*{Differentiable Immersed Boundary Solver}
The immersed boundary method (IB) is widely used for simulating complex fluid flows, especially when dealing with fluid-structure interactions  \cite{shaqfeh_IBM,IBM_Second_order,Ulhmann_IBM}. It was originally developed to address problems where the geometry of the immersed objects is not aligned with the grid used for solving the fluid flow equations \cite{peskin_2002}. The re-meshing independence of IB when applying the no-slip boundary condition to moving or deforming objects makes it well-suited for shape optimization problems and differentiable programming.

In the IB method, objects are depicted in a Lagrangian framework. Their positions and characteristics are independently tracked, separately from the stationary Eulerian grid employed to numerically solve fluid flow, typically through techniques like finite differences or finite volumes. At each time step, forces are computed at the fluid-object interface and are then used to influence both the fluid velocity field and the movement or deformation of the immersed object. Significantly, the immutability of both the Lagrangian and Eulerian grids greatly streamlines the GPU implementation of the method \cite{IBM_GPU}.

Various IB implementations have been developed, with different approaches for coupling forces on immersed objects with fluid flow modifications \cite{IBM_annual,taira,ibm_direct,wang_IBM}. We have developed a differentiable variant of the direct forcing IB, originally formulated by Uhlmann for rigid bodies \cite{Ulhmann_IBM}, and integrated this version into the differentiable JAX-CFD package \cite{jaxcfd}, developed for solving Navier-Stokes equations via the projection method. The details of the algorithm are outlined in the supplemental information. We chose the kernel for convolving the force from the Lagrangian grid to the fluid domain as an exponential function $\delta(x) = e^{-x/h}$ where $h$ is the step size, and the number of Lagrangian grid points is selected such that the distance between them is close to the grid size of the Eulerian mesh. The remainder of the implementation closely follows Uhlmann's algorithm.  

We validated the differentiable solver by comparing (see Figure S1) the time-dependent drag force over an ellipse rotating and translating following a sinusoidal wave function with published numerical results \cite{eldredge,wang}. We calculated the drag force by integrating the total force the fluid exerts on the particle. 

The remainder of this paper explores various high-dimensional optimization problems solved using this library. The range of problems explored underscores the versatility and ease of implementation of Automatic Differentiation (AD) compared to the adjoint-based method, highlighting its effectiveness in tackling complex flow problems. In each problem,  we carry out the optimization  using the interior point optimization with the canonical library Ipopt \cite{ipopt}. The gradients of the loss function and the constraints are evaluated using the backward-mode of automatic differentiation when the number of degrees of freedom is lower than the number of constraints, and the forward-mode was used when the number of constraints is higher.

\begin{figure*}
\centering
\includegraphics[scale=0.35]{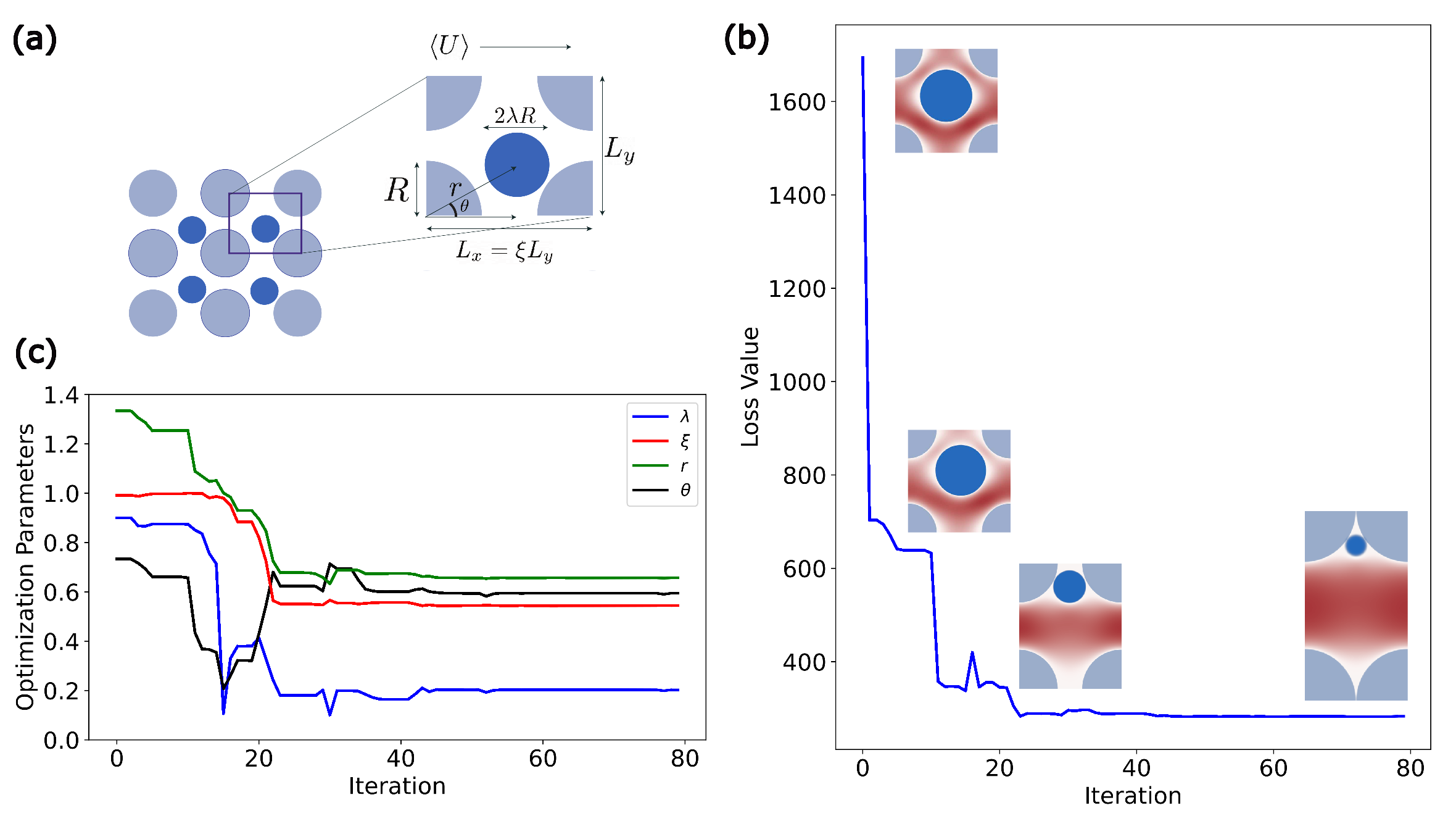}
\caption{\label{fig:porous_loss} {\bf Optimal Porous Media for flow rate} a) Sketch of the periodic porous medium. We optimize over the position of the central (dark blue) particle, the ratio of sphere radii ($\lambda$), and the aspect ratio $\xi$. b) The optimization trajectory of the loss function. The pictures show how the network topology  changes at chosen optimization iterations. c) The value of control parameters during the optimization iterations. After about 50 iterations, each parameter converges to an optima }
\end{figure*}

\begin{figure}[t]
\centering
\includegraphics[scale=0.25]{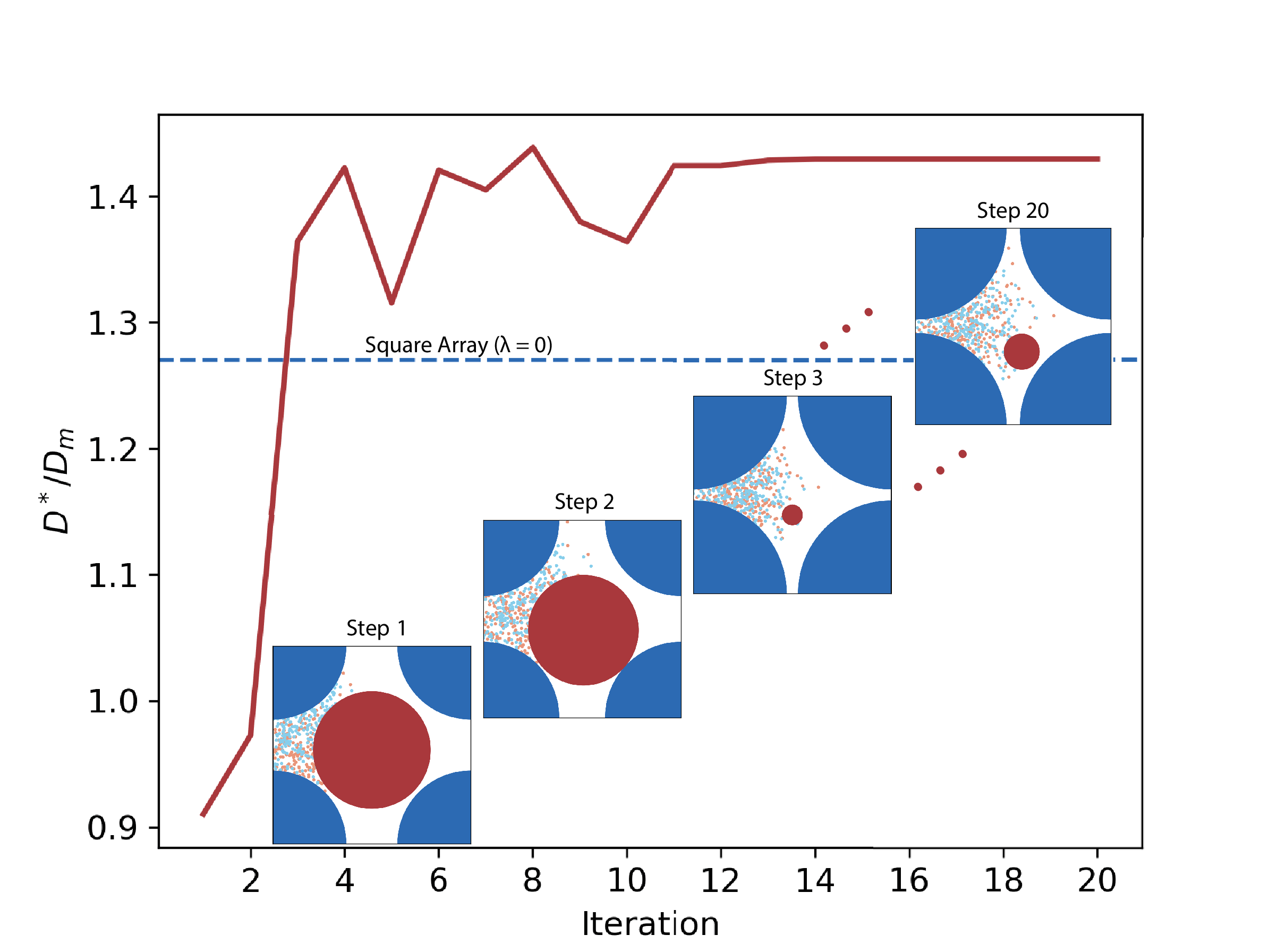}
\caption{\label{fig:disp_coeff} {\bf Optimal Porous Media for Dispersion} Optimization trajectory of the scaled dispersion coefficient ($D^*/D_m$) for tracer particles in the flow.
We initialize the blue (orange) particles  in the top (bottom) half of the gap at $x=0$.
The figure shows porous medium geometry and the positions of two colors of tracer particles after 100,000 time steps at different optimization iterations. The dashed blue line represents the scaled dispersion coefficient for the simple square array ($\lambda=0$).  }
\end{figure}

\section*{Flow in Porous Media}
Optimizing the topology of porous media to control flow behavior is a significant challenge with numerous real-world applications. Pioneering work in this area derived an analytical solution to predict the pressure drop in a zero Reynolds number flow passing Bravais lattices of cylindrical posts \cite{acrivos}.  Others have studied the influence of polydispersity \cite{Edwards}, fluid rheology \cite{power_law}, and Reynolds number \cite{koch_ladd_1997} on the flow behavior. However, there have been limited attempts to optimize the arrangement of arrays to minimize drag \cite{cylinderopt} in any of these situations. For instance, \cite{optimizationdrag} optimized the arrangement of a number of cylinders to minimize wake formations, showing the significance of relative position on the flow behavior within a porous medium. We are not aware of other substantial investigations carried out to systematically enhance the topology of such porous media.

Here, we optimize the packing of a bidisperse system of periodically arranged circular rods to maximize the flow rate along the direction of applied pressure gradient for a given solid packing fraction. We consider a pressure-driven flow of an incompressible Newtonian fluid through a periodic structure of a bidisperse assembly of circular cylinders, as depicted in Figure  \ref{fig:porous_loss}a. The unit cell of the periodic array is characterized by its aspect ratio $\xi$ which ranges between $0$ and $1$, the cylinders' radius ratio, $0 \leq \lambda \leq 1$, and the relative position of the dissimilar cylinders characterized by $r$ and $\theta$. Here, $r$ is the distance between the center of the cylinders, and $\theta$ is the angle between the position vector of the interior cylinder (dark blue circle in figure \ref{fig:porous_loss}a) and the lattice (horizontal) $x$-axis. When the values of $\lambda = 1$, $\xi=1$, $r = \sqrt{2}L_y$, and $\theta = 45^{\circ}$, the array corresponds to a monodisperse staggered square while a simple square array is obtained when $\lambda = 0$, $\xi=1$. The Reynolds number in this problem is denoted as $Re_M = \left< U \right> D_M/\nu$ where $\left< U \right>$ is the average fluid velocity along the unit cell's $x$-axis, $D_M$ is the monodisperse particle diameter and is equal to $D_{M}=L_y \sqrt{\frac{2\xi\phi}{\pi}}$ and $\nu$ is the kinematic viscosity. In the problem, the Reynolds number is set to be equal to $10^{-4}$. Because the particles are stationary, the force coupling can be simplified by representing the forces the particles exert on the fluid as a drag field following Brinkman's penalty method \cite{brinkman1,brinkman2}.  

For a solid volume fraction, $\phi = 0.44$, we optimize these lattice parameters to find the optimum array topology that minimizes the scaled macroscopic pressure drop, $ \Delta P L_x^{2}  / \langle U \rangle \mu \xi^{2}$, along the flow direction. We constrain the optimization by enforcing a non-overlapping condition, yielding a constrained non-linear optimization problem.  Figure \ref{fig:porous_loss}b shows the optimization trajectory of the loss function while Figure \ref{fig:porous_loss}c shows the trajectory of the optimization parameters including the unit cell aspect ratio, $\xi$, the particle size ratio, $\lambda$, and the position parameters for the interior cylinders. Furthermore, the images offer visual insight into the dynamic transformation of the porous medium's topology at various optimization iterations. The optimization process unfolds as the staggered square array evolves into a rectangular array configuration, strategically positioning the interior cylinder between corner ones. This transition effectively increases the gap thickness and reduces the flow path tortuosity. This outcome is aligned with the expectations derived from lubrication theory, where the pressure drop is proportional to the cube of the gap thickness.

\section*{Dispersion in Porous Media}

We next consider dispersion of tracers in porous media, where evaluation of the loss function and its gradient relies on statistical metrics derived from the trajectories of stochastic tracer particles. This scenario presents a complex optimization problem since the computed gradients are noisy due to the randomness of the particle trajectories. Finding optimal solutions for dispersion in porous media not only poses a theoretical challenge but also holds practical significance in several fields such as groundwater remediation, oil reservoir management, and pollutant transport.

\begin{figure*}
\centering
\includegraphics[scale=0.3]{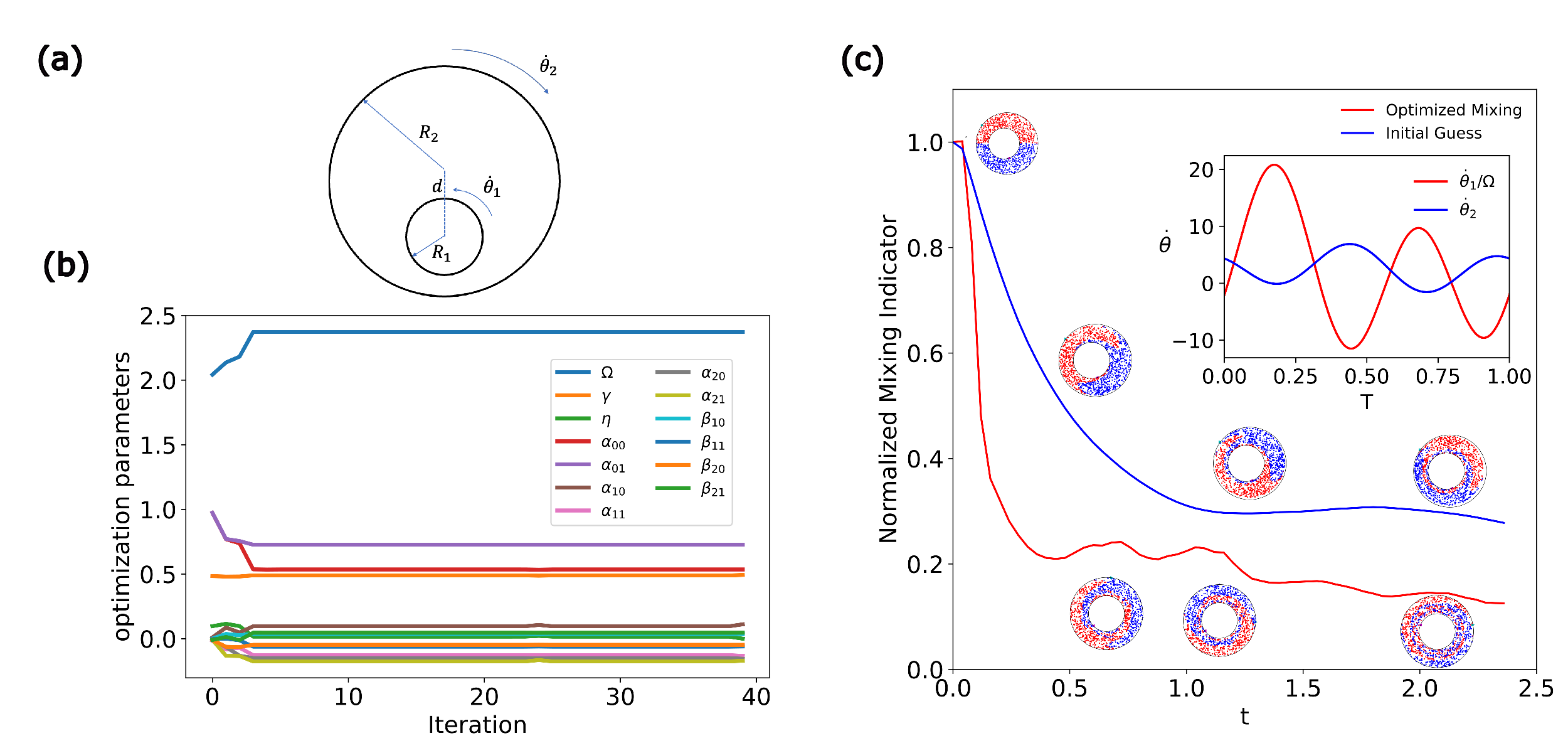}
\caption{\label{fig:mixing_setup} {\bf Chaotic mixing in a journal bearing} (a) Mixing problem set-up illustration, with two non-concentric cylinders whose centers are displaced by $d$ of different radii $R_1,R_2$ being rotated with different angular velocities $\dot\theta_{1,2}$. The optimization parameters not only include geometrical parameters but also Fourier coefficients of the forcing (5 for each cylinder). (b) Trajectory of the design parameters after a number of optimization iterations. After a small number of iterations they settle to an optimum (c) The transient profile of the mixing indicator. The blue curve corresponds to the case of defined parameters while the red curve corresponds to the case using optimized parameters. The pictures illustrate the mixing at various points of time. The inset figure shows the optimum rotation velocity of the outer and inner cylinders. Here, $T=\tau f$ where $f$ is the rotation frequency.}
\end{figure*}

Dispersion within two-dimensional periodic porous media has been a subject of extensive exploration \cite{brenner1,koch_dispersion,2d_dispersion}. However, there has been a notable absence of research dedicated to the inverse problem – that is, the modification of porous media topology to attain a targeted dispersion tensor. Such an inquiry, particularly when considering transport of tracer particles modeled through Brownian dynamics, represents an uncharted territory in this field of study. Without loss of generality in the method to target a dispersion tensor, we consider optimizing the axial dispersion coefficient $D^*$ given by:
\begin{equation}\label{eq:langevin}
   D^*=\frac{1}{2}\lim_{t\to\infty}\frac{d\sigma^2(t)}{dt} 
\end{equation}
Here, $\sigma^2=\big\langle \left(\mathbf{X}_1^i(t)-\left\langle \mathbf{X}_1^i(t)\right\rangle\right)^2\big\rangle$ where $\mathbf{X}_1^i(t)$ is the $x$-component of the position vector of tracer, $i$.   Note that this dispersion coefficient is averaged over the entire flow domain, so it implicitly accounts for stagnant regions in the flow.

 To compute the dispersion coefficient, we compute the steady fluid velocity field, $\mathbf{u}$, for a given porous medium geometry following the same setup as described in the previous section. We then use
 the Langevin equations to model the trajectory of the test particles  \begin{equation}\label{eq:potential} \frac{d\mathbf{X}^i}{dt}=\mathbf{u} +\nabla\sum_{j=1}^NU_j(\mathbf{X}^i, r_j)+\sqrt{2D_m \boldsymbol{\xi}(t)}.\end{equation} Here, $U_j$ represents the repulsive potential of the $j_\text{th}$ cylindrical post, which has radius $r_j$. We use a Morse repulsive potential to model hard particles \cite{morse} (see SI for details). Here, $D_m$ is the molecular diffusion coefficient of the tracer particles, and
 $\boldsymbol{\xi}$ is white noise drawn from a Gaussian distribution. We control the dimensionless
 Péclet number, defined as $Pe= \langle U\rangle R_M/D_m $ where $R_M$ is the monodisperse particle radius of the porous media given by $R_m = L_y/2 \sqrt{\frac{2\xi\phi}{\pi}}$.  We integrate Equation \ref{eq:potential} with a differentiable brownian dynamics solver \citep{jaxmd2020}. 

 Computing the dispersion coefficient by tracking Lagrangian tracer particles is a multiscale problem in that we need to resolve the diffusion of particles across streamlines at small timescales, while also running the simulations long enough that the mean square displacement of the particles reaches the diffusive regime. For the optimization results presented here, we simulate particle trajectories with a step size of $dt=0.001$ for a total of 1.5 million steps. To adequately sample the unit cell and reduce noise in the computation of the dispersion coefficient, we use 1,200 tracer particles initialized in a vertical line at $x=0$ in the gap between the corner posts. For rapidly computating the dispersion coefficient, we developed a differentiable implementation of an autocorrelation-based algorithm for computing the lagged mean squared displacement curves from particle trajectories \citep{fft-msd}.

 Figure \ref{fig:disp_coeff} shows the optimization trajectory of the loss function, defined as $D^*/D_m$ when the packing density, $\phi$, is equal to $0.7$ and the Peclet number is set to $Pe=1.0$. In this problem, we set the unit cell aspect ratio, $\xi$ to be equal to $1$ while we optimize the relative position of the interior post and the radii ratio between the interior cylinder and the corner ones, initialized with the particle located at the center of the unit cell and a particle radius ratio of $\lambda=0.8$. The snapshots show the dynamic transformation of the topology of the porous media to maximize the scaled dispersion coefficient. As can be seen in the figure, the solver was able to modify the initial geometry of a nearly monodisperse staggered square array to increase the dispersion by more than 40$\%$. As a comparison, the scaled dispersion coefficient for a simple square array system, i.e. $\lambda=0$, is 1.27. Thus, our optimization procedure identifies a geometry with a higher dispersion coefficient than typically studied Bravais lattices. It is worth mentioning that for the initial geometry used in the optimization, the dispersion coefficient is lower than the molecular diffusion constant, owing to large stagnant regions in the flow, indicating that the iterative optimization of the average dispersion coefficient shrinks  the size of stagnant flow regions. 

\section*{Mixing in Journal Bearing}\label{sec:mixing}
Now, let us consider an inverse flow problem involving unsteady flows. Achieving efficient and controlled mixing in low Reynolds number flows, particularly within the Stokes flow regime, represents a critical challenge with widespread applications in microfluidics, drug delivery, and chemical reactions. The utilization of rotating cylinders as mixers has been shown to induce chaotic mixing even at very low Reynolds numbers when the generated flow is time-dependent \cite{umass}. 

There is a rich body of research characterizing the chaotic nature of the generating flows \cite{aref_1984,aref_review}, though only few studies have considered optimizing the protocols that govern the rotation of these cylinders to increase the efficiency of mixing. Here, we optimize the configuration and protocol of rotating eccentric cylinders, shown in figure \ref{fig:mixing_setup}a, to increase the efficiency of mixing. 
We define the degree of mixing as the average distance between two groups of different colored fluid particles that are initially separated apart, a definition proposed  by \citep{stone_mixing} to estimate the mixing in a rotating droplet. The optimization parameters we consider are the diameter ratio of the two cylinders, $r$, the eccentricity, $\eta$, defined as the distance between the two cylinders normalized by the radius of the outer one, and the Fourier series coefficients, $a_{in}$ and $b_{in}$ that describe the time dependent rotation of cylinder $i$ as:
\begin{equation}
    \theta_i (t) = \theta_0 + \sum_{n=0}^N a_{in} {\rm{cos}}(2\pi f t+\phi) + b_{in} {\rm{sin}}(2\pi f t+\phi_i),
\end{equation}
with $N$ the maximum number of modes used to describe the oscillations.

The trajectory of the fluid particles is solved using:
\begin{equation}
    \frac{d\mathbf{X}_i}{dt} = \mathbf{u}(t),
\end{equation}
where  $\mathbf{X}_i$ is the position of fluid particle $i$ and  $ \mathbf{u}$ is the time-dependent fluid velocity solved using the IB solver.

Figure \ref{fig:mixing_setup}b shows the convergence trajectory of the optimization parameters, where we have set $N=5$. The efficiency of mixing is calculated after two periods of rotations. Figure \ref{fig:mixing_setup}c displays the normalized transient mixing indicator using both the optimized design and the initial guess solution for the journal bearing. The parameters for the base case are set as follows: $\Omega=2$, $a_{in}=b_{in}=0$ for $i>0$, $a_{01}=0.2$, $a_{00}=0.2$, and $b_0$ and $\phi=0$. The solid blue curve represents the transient profile of the normalized mixing indicator before optimization, with normalization performed using the initial average distance between the blue and red test particles. The images demonstrate the mixing process overtime for both the base case and the case with the optimized protocol. The inset figure illustrates the optimal kinematics of the rotating cylinders during a rotation period.

 Interestingly, our findings reveal that counter-rotation of the cylinders is not necessary to achieve maximum mixing; instead, a phase shift between their rotation periods suffices. In our exploration of this problem, we conducted various optimization iterations with distinct initial guesses, each leading to significantly different optimal solutions, all with similar mixing efficiencies (figure not shown). These solutions exhibit significantly different values of $\Omega$ and $\eta$, indicating the degeneracy of the optimal mixing problem. Chaotic mixing in 2D is readily attainable when the fluid flow is time-dependent, resulting in multiple, diverse optimal solutions.

\begin{figure*}[t]
\centering
\includegraphics[scale=0.3]{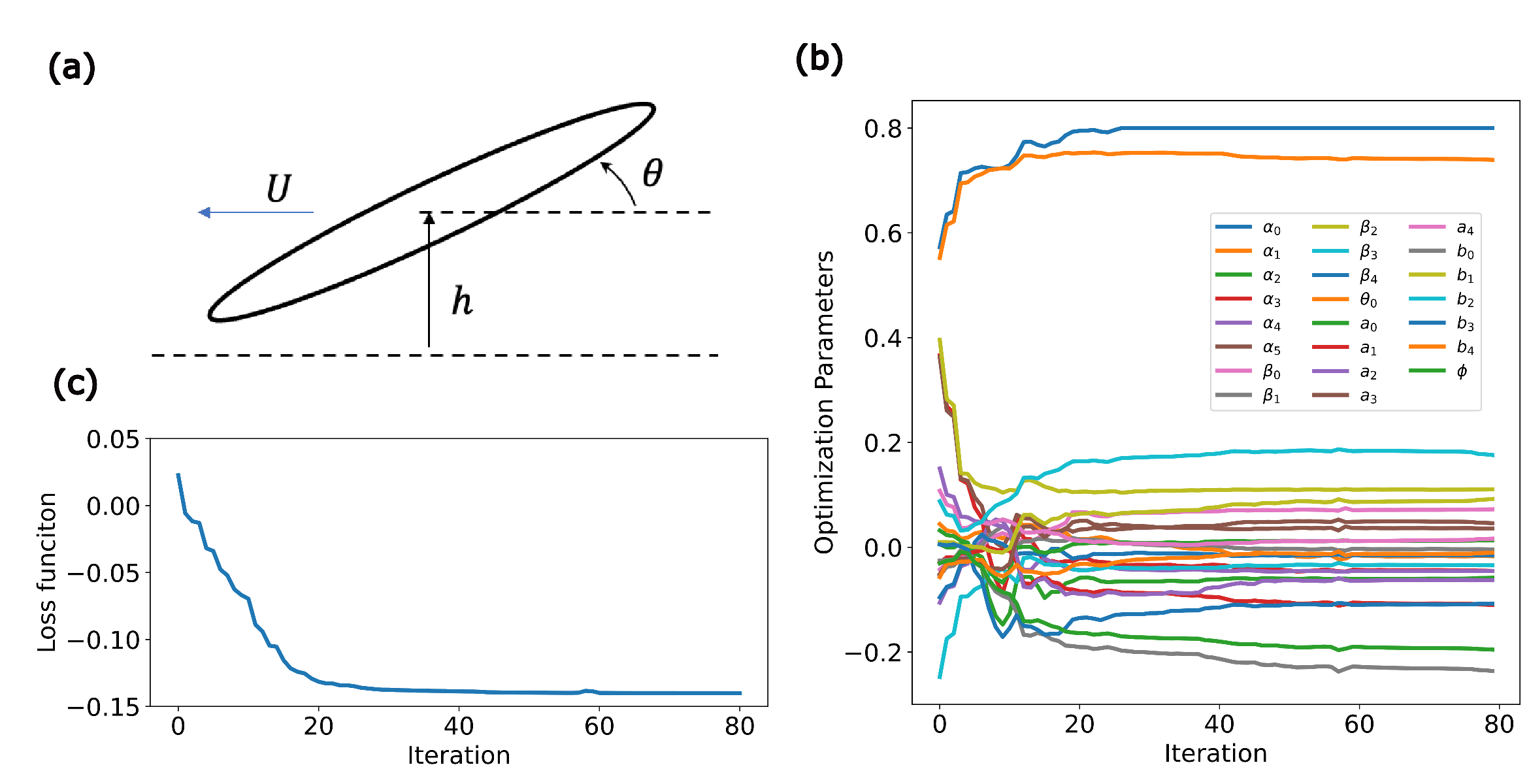}
\caption{\label{fig:swim_sketch} {\bf Optimal Swimmer} (a) Sketch of the swimming problem, where a swimmer with an elliptical cross section moves with velocity $U$, using a stroke $h(t)$ and $\theta(t)$. Both $h(t),\theta(t)$ are parameterized with $20$ parameters. (b) Convergence of these design parameters over 80 iterations for the case $Re=1000$. (c) Loss function (swimming efficiency, the ratio of generated thrust to power input) as a function of iteration. The efficiency dropped more than twofold during the optimization}
\end{figure*}

\section*{Swimming Efficiency}\label{sec:siwmmer}
In the previous problems, we focused on flows at low Reynolds numbers. The implementation of AD in differentiating over a full Navier-Stokes equation solver also enables us to design high Reynolds number flows. To illustrate this capability, we consider the optimization of flapping motion, a fundamental mode of locomotion seen in various organisms, including insects, birds, and fish. While pioneering work has shed light on some thrust generation mechanisms, a comprehensive understanding of how to maximize swimming efficiency remains incomplete. Classical studies have highlighted the importance of factors such as vorticity fields generated during simple kinematic motions, where objects oscillate vertically (heaving motion) and flap periodically (pitching motion) with a single frequency, leading to the generation of a reversed von Kármán vortex street (see ref. \cite{swim_review} for a comprehensive review). Focusing on single frequency swimming, several experimental and theoretical studies have determined an optimum frequency value for different types of swimmers and the role of leading-edge vortices in thrust generation \cite{optSt_nature,nature_flapping}. Investigating vorticity profiles for more efficient swimming offers valuable insights into this intricate problem, making it an ideal case for leveraging automatic differentiation to find these fields.

Here, we optimize the hovering and pitching kinematics of a 2D ellipsoid airfoil moving forward at a speed denoted as $U$ in an unbounded fluid, as illustrated in Figure \ref{fig:swim_sketch}a. To increase the complexity of the problem, we consider multiple frequencies to describe the hovering motion, represented by a Fourier series:
\begin{equation}
    h(t) = \sum_{m=0}^M \alpha_m {\rm{cos}}(2\pi f t) + \sum_{n=0}^N \beta_n {\rm{sin}}(2\pi ft)
\end{equation}
and, similarly, with  pitching motion  described by:
\begin{equation}
    \theta (t) = \theta_0 + \sum_{m=0}^M a_n {\rm{cos}}(2\pi f t+\phi) + \sum_{n=0}^N b_n {\rm{sin}}(2\pi f t+\phi)
\end{equation}
We define the efficiency of swimming  as the ratio of the generated thrust to the power input represented by the lift and moment of rotation,  given by:
\begin{equation} \label{eq:loss_swim}
    \mathcal{L}= \frac{\int_0^T F_x U dt}{\int_0^T F_y \dot{h} + M\dot{\theta} dt }
\end{equation}
where $F_x$ is the generated thrust, $F_y$ is the lift force while $M$ is the moment of rotation. $\dot{x}$ represents the time derivative. 
We maximized the efficiency of a swimming ellipse with a chord length of $c=1.0$,  aspect ratio of $8.33$, and  Reynolds number of $R_e = Uc\rho/\mu=1000$,  where $\rho$ and $\mu$ are the fluid's density and viscosity, respectively. Figure \ref{fig:swim_sketch}b shows the trajectory of the optimization parameters used to describe the oscillatory translation and rotation of the ellipse after multiple optimization iterations. Figure \ref{fig:swim_sketch}c shows the trajectory of the loss function, propulsion efficiency, after multiple iterations. 

\begin{figure*}[t]
\centering
\includegraphics[scale=0.3]{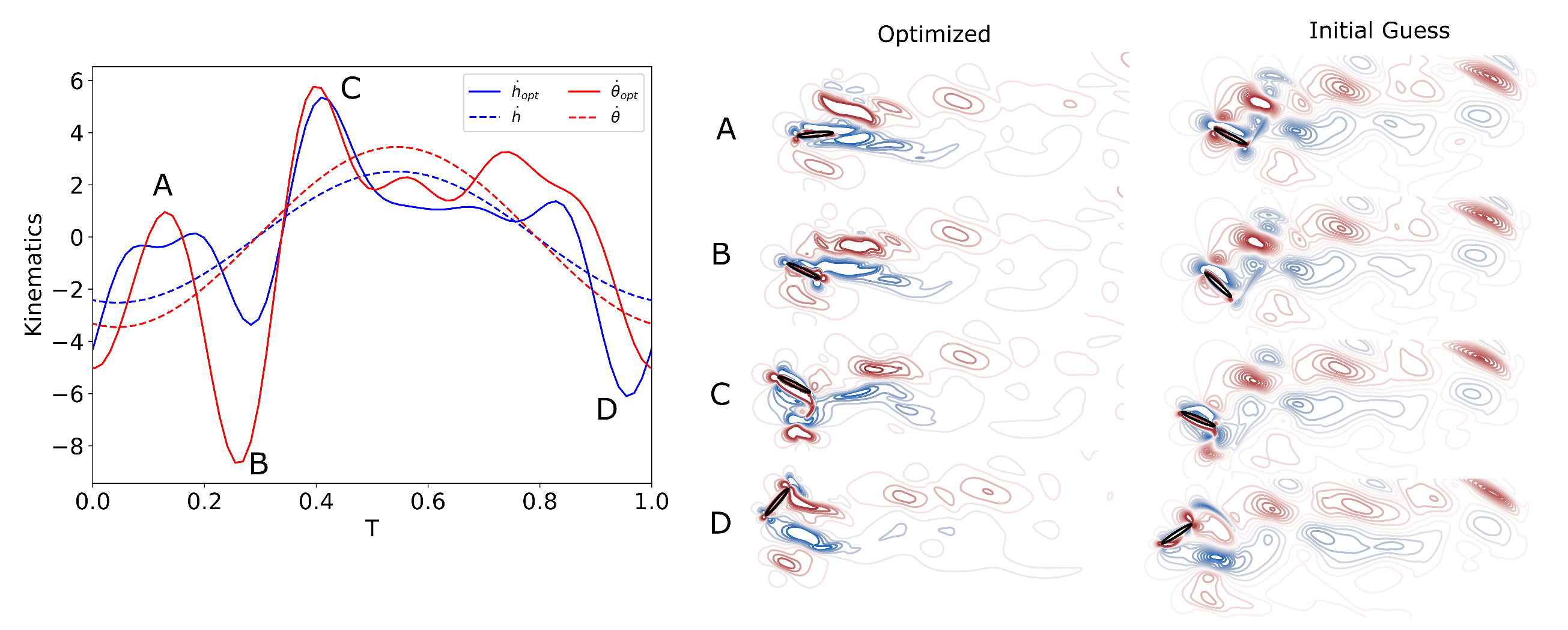}
\caption{\label{fig:optimum_kin_swim} A comparison of the optimum periodic heaving and pitching motions with the single frequency motion used as an initial guess. The solid blue and red curves represent the optimum heaving and pitching motion, respectively. The dashed curves show the single frequency kinematics. $T=\tau f$ where $f$ is the frequency. The images on the right side show the vorticity contour resulting from the optimized motions at various points of time and that resulting from a simple single frequency motion.}
\end{figure*}

Figure \ref{fig:optimum_kin_swim} shows the optimum kinematics over a single stroke period compared with the single frequency kinematics that were used as an initial guess. The solid blue and dashed blue curves represent the optimum vertical velocity of the swimmer, $\dot{h}_{opt}$, and the vertical velocity of the single frequency kinematics, $\dot{h}$, respectively. The solid red and dashed red curves represent the optimum rotational rate, $\dot{\theta}_{opt}$, and the rotational rate of the single frequency case, $\dot{\theta}$, respectively. One interesting feature of the solution is that there is not a significant phase shift between the heaving and pitching motion as the optimum value of $\phi$  converges to 0.1. 

Interestingly, we found that while the optimum vorticity field generates the expected reversed von Kármán street, the pattern shows that unlike the single frequency locomotion where the two-counter-rotating vortices are at an angle as shown in Figure \ref{fig:optimum_kin_swim}c, the optimized swimmer tends to generate almost vertically aligned counter-rotating vortices. The swimmer generates vortices and then reinforces such vortices by synchronizing the strikes at the center of the generated vortex. This is evident by looking at the motion of the swimmer from case D to case A in the optimized vorticity profile plot, Figure \ref{fig:optimum_kin_swim}b. The swimmer strikes in the same position as the blue vortex. Finally, we observe that the generated vortices are large and span the whole body of the swimmer integrating the leading and the trailing edge vortices.   

\section*{Discussion and Conclusion}\label{sec:siwmmer}
There has been much discussion about the application of machine learning to the sciences, such as fluid mechanics\cite{brenner2019perspective}. We feel strongly that the emphasis on learning or emulating properties of flows misses a key point, which is that the underlying {\sl causes} of the technological revolution that have given rise to the remarkable advances in machine learning offer the potential for solving   classical fluid mechanics problems in a different way without using any machine learning {\sl per se}, but simply by using the computational infrastructure. The core of this infrastructure is 
automatic differentiation, a highly efficient technique for computing derivatives of solution trajectories of partial differential equations. In this paper, we implement the immerse boundary method in JAX and demonstrate that the resulting code offers a remarkably flexible interface for solving a wide range of flow optimization problems.  Key elements of the method are that it allows for solving optimization problems with large numbers of parameters, while optimizing complicated cost functions. The flexibility of the method makes it possible to rapidly experiment with different approaches and formulations, finding versions that are easier or more intuitive to optimize. This contrasts with the markedly less flexible classical use of adjoint methods.  As an example, we demonstrated that it is easily possible to formulate an optimization problem that simultaneously optimizes a hybrid Eulerian/Lagrangian calculation for chaotic mixing. The computation graph of the problem does not structurally change by stacking a Lagrangian calculation for the passive tracers on top of an Eulerian calculation for the underlying flow -- even the stochasticity of particle trajectories is easily dealt with by computing gradients with respect to expectations of stochastic variables.

We used this streamlined optimization methodology to tackle a multitude of canonical inverse flow problems, ranging from swimming at high Reynolds numbers to differentiating through stochastic Lagrangian trajectories. 
The methodology scales favorably with the number of input parameters, enabling  the  design  of  increasingly complex flows. The fluid flow in porous media problem showcased the ease with which boundary conditions can be modified to target specific flows. The swimming problem revealed the rich possibilities of vortex fields that maximize the swimming efficiency opening the door to developing better understanding of the locomotion of various organisms. The ability to differentiate through transient profiles enabled us to invert unsteady flows to optimize the mixing of journal bearing.  While we focus in this paper on solving two-dimensional problems, it is important to emphasize that the solver can be used for three-dimensional scenarios as well. There are technical challenges that arise,  though  we are optimistic that these can be surmounted, either by surrogate modeling, distributing the solver across multiple GPUs, using the implicit differentiation theorem \cite{implicit_diff}, or other approaches.

A major opportunity for future work is to simultaneously improve {\sl both} the ability to model complex systems while at the same time solving inverse problems. A recent paper demonstrates how physical laws can be extracted from pixel-by-pixel particle image  data \cite{bazant23}; such an approach can be easily implemented using differentiable models.  Another illustration is a recent paper\cite{kochkov2023neural}, which demonstrated a state-of-the-art general circulation model (GCM) whereby the physics parameterizations are {\sl learned} from fitting to global weather data for 3-5 day forecasts.  This learning is possible because the parameterization can be formulated as an optimization problem since the underlying GCM is differentiable. Fluid mechanics is a discipline filled with parameterizations, which occur when we are trying to solve equations in a regime where either the models are not accurate or we do not have sufficient resolution. Examples range from rheology or dynamics of polymer solutions, to heat transfer at high Rayleigh and Reynolds numbers, to high Reynolds number boundary layer separation. There is tremendous potential to use differentiable codes to first improve an underlying model and then to optimize the model to target desired behaviors and discover novel ones. The developed code can be found in the following repository \url{https://github.com/hashimmg/jax_IB/} with a few examples demonstrating various problem setups.  

The authors would like to thank David Weitz for valuable discussions. This work was supported by  the Office of Naval Research through grant numbers ONR N00014-17-1-3029, ONR N00014-23-1-2654 and National Science Foundation through DMR-2011754 and the NSF AI Institute of Dynamic Systems (2112085). M.G.H was supported by Saudi Aramco.

\bibliography{pnas-sample}
\end{document}